\documentclass[letterpaper,twocolumn,10pt]{article}
\usepackage{usenix2021}
\usepackage{xspace}
\usepackage{graphicx}

\usepackage[capitalise,nameinlink,noabbrev]{cleveref}     %

\newcommand{\name}{TerraWatt\xspace}
\newcommand{\containername}{SunDrop\xspace}
\newcommand{\containernames}{SunDrops\xspace}

\begin{document}
\date{}

\title{\Large \bf \name: Sustaining Sustainable Computing of Containers in Containers}

\author{
{\rm Jennifer Switzer}\\
UC San Diego
\and
{\rm Rob McGuinness}\\
UC San Diego
\and
{\rm Pat Pannuto}\\
UC San Diego
\and
{\rm George Porter}\\
UC San Diego
\and
{\rm Aaron Schulman}\\
UC San Diego
\and
{\rm Barath Raghavan}\\
USC
} %

\maketitle

\begin{abstract}
Each day the world inches closer to a climate catastrophe \emph{and} a sustainability revolution. To avoid the former and achieve the latter we must transform our use of energy.
Surprisingly, today's growing problem is that there is too much wind and solar power generation at the wrong times and in the wrong places.

We argue for the construction of \name: a geographically-distributed, large-scale, zero-carbon compute infrastructure using renewable energy and older hardware.
Delivering zero-carbon compute for general cloud workloads is challenging due to spatiotemporal power variability. We describe the systems challenges in using intermittent renewable power at scale to fuel such an older, decentralized compute infrastructure.
\end{abstract}

\section{Introduction}

To reduce the carbon impact of datacenters, operators have pledged to rely on renewable energy. Sometimes they have placed datacenters near cheap, renewable generation, such as Google's facility at The Dalles, Oregon. They also time-shift workloads to use periodically-abundant renewable power~\cite{google_zero_carbon}. 

However, few cloud service providers have demonstrated that they can achieve zero-carbon compute that uses only renewable energy and hardware whose embodied energy---the energy used in mining and manufacturing---is zero or near zero.
Zero-carbon compute avoids the waste of (1) renewable energy that is thrown away or sold at a negative price and (2) embodied energy of hardware that is thrown away when machines are retired. It is this challenge we tackle in this paper.

The electric grid is often unable to collect and store energy or transport it to where it could be used. These issues are both fundamental (e.g., intermittent renewable sources, losses due to transport inefficiency, lack of grid interconnection) and practical (e.g., high energy storage costs). While it is hard to move remote power to data centers, moving data and compute near renewable generation is feasible. 
At the same time, compute, storage, and networking gear within data centers have a fast hardware refresh cycle.
Since the vast majority of a computer's carbon footprint stems from its manufacture~\cite{c02-server}, short lifespans waste significant embodied energy~\cite{emergy11}. 

Researchers~\cite{chien1,chien2,chien-extendinglife} and industry~\cite{soluna,lancium} have found that
bulk-compute workloads are the easiest to run on such zero-carbon compute.
Such workloads have loose deadlines and can be paused, which allows them to move toward renewable power and stop when it becomes unavailable. Although bulk workloads are not particularly efficient on older compute, they make use of hardware that would otherwise go to waste.
However, prior work has considered the hardware and software abstractions neither to support zero-carbon compute for \emph{general-purpose} cloud compute nor to tolerate the data movement and intermittent power inherent in this context.

Carrying these ideas to fruition gives license to a seemingly-fantastical dream: that there is no fundamental barrier to building (distributed) hyper-scale data centers that have zero or near-zero carbon footprint. Since there is often surplus power available \emph{somewhere}, such data centers could provide high availability (e.g., the wind is blowing in the US Midwest at night while the sun is shining in the US West during the day).

To realize this dream we must rethink the systems and networking abstractions that underpin the infrastructure of this new type of cloud platform. In this paper we imagine the construction and deployment of \name, a geographically-distributed, zero-carbon compute infrastructure, based on the trends highlighted above. This infrastructure is instantiated as individual shipping containers called \containernames with predominantly old hardware. The three central challenges in designing and building \name are (1) designing abstractions and infrastructure for distributed, intermittently-powered compute that expose some but not all of the vagaries of the underlying power availability, (2) addressing the systems challenges in using legacy compute infrastructure to still service meaningful workloads, and (3) designing a framework and metrics for evaluating the energy and carbon footprint of \name at macro-scale and individual tasks run on it at micro-scale.

\section{Impacts and opportunities}

We begin with the current limits and opportunities of renewable energy sources. We then describe how we can use curtailed and negative-priced power with recycled hardware to build zero-carbon compute infrastructure.

\begin{figure}[t]
\centering
    \includegraphics[width=\linewidth]{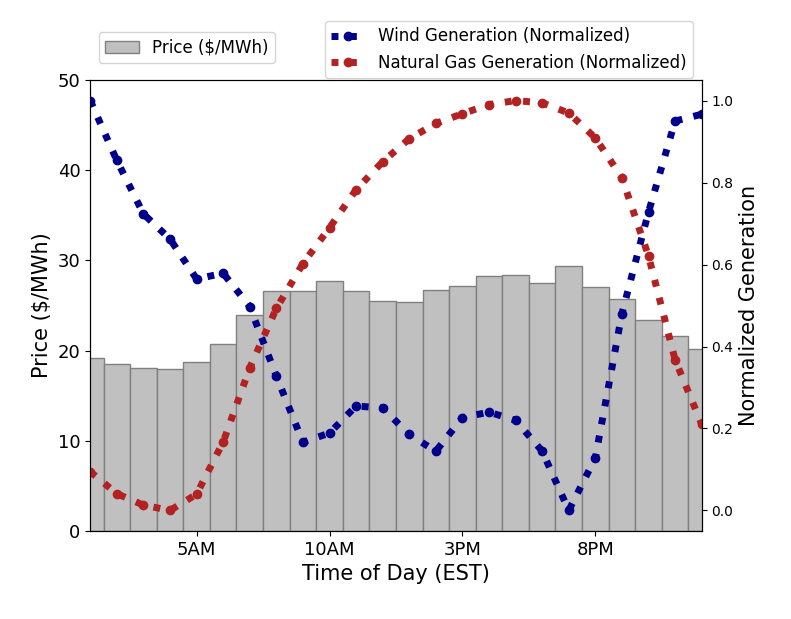}
    \vskip -1em
    \caption{Gas-fired plants can be adjusted to meet demand (e.g., produce more when prices are higher); wind energy is weather-dependent and often not in phase with demand.}
    \label{f:wind-motivation}
    \vskip -1em
\end{figure}

\subsection{Limits of the grid}

Renewables became cheaper than coal in 2018~\cite{lazard2018levelized}.
In 2019, 40\% of new U.S. generation capacity came from solar~\cite{lotsofsolar}.
However, adding more capacity alone isn't enough.

Renewables' \emph{intermittency} is their Achilles heel.  Solar and wind, in particular, fluctuate with weather and solar irradiance~\cite{miso, caiso} and are not in phase with one another. This  puts them at a disadvantage compared to fossil fuel-based energy sources, which adjust production to meet demand, as shown in \cref{f:wind-motivation}. 

\subsubsection{Opportunity Power}
Renewables have periods of underproduction, during which not enough renewable energy is available to meet demands, and periods of overproduction, during which \emph{too much} renewable energy is available, to the point that the excess must be discarded (curtailed) or sold at a negative price point~\cite{chien1}. Negative-price power and curtailed power are together referred to as \emph{opportunity power}~\cite{miso}. 

Opportunity power in the U.S. is significant, growing, and often available. North America's two largest ISOs---the California Independent System Operator (CAISO) and the Midcontinent Independent System Operator (MISO)---together produced between 7--20\,TWh of opportunity power in 2017~\cite{miso, caiso}. Opportunity power is available somewhere in MISO 99\% of the time---since prices are location-dependent---and often in intervals of >100 hours~\cite{miso}. In CAISO, some solar generators experience 3.3 hours of opportunity power per day~\cite{caiso}.

As solar and wind generation grow, so too will the amount of curtailed and negative-priced power. Indeed, CAISO estimates their compound annual growth rate of opportunity power to be 40\%~\cite{caiso}.
Assuming 1.5\,TWh of opportunity power in CAISO in 2017 (a conservative estimate) and a constant growth rate, CAISO alone could provide 22\,TWh of opportunity power by 2025, enough to power the city of Los~Angeles.

\subsubsection{Storage limits}
Energy storage is expensive: grid-scale lithium-ion batteries cost approximately \$209 per kWh~\cite{batteryprices}, not including installation costs.
A basic analysis of CAISO and MISO, assuming 1.5\,TWh/year and 6\,TWh/year of opportunity power, respectively, yields a conservative estimate of \$35\,million to add one hour of storage to CAISO and \$140\,million to add one hour of storage to MISO.
A more advanced analysis by MISO suggests that adding grid-level storage provides diminishing returns, and that adding 50 hours of grid-level storage to MISO would cost \$50-\$400M \textit{per wind generation site}~\cite{miso}, a price tag on par with the cost of the wind turbines themselves.

\subsubsection{Transport Limits and Economics}
Another simple response to excess renewable power is to deliver it to existing datacenter deployments. Unfortunately, the cost to deliver power grows non-linearly with distance, with transmission lines alone varying widely due to land and construction costs, often costing anywhere from \$2,500 per MW-mile to \$16,000 per MW-mile\cite{andrade2016transmission}. 
Other work~\cite{fares2017transmissioncost} finds further combined transmission, distribution, and administrative costs for power over time.

\subsubsection{Demand shifting and data centers}

Renewable energy suffers from a supply-demand mismatch, but shifting the supply in time and space is infeasible. Many renewable energy proponents have recognized the need for so-called flexible loads, which adapt in real-time to the carbon intensity of the grid~\cite{flex1, flex2,demandbal}. Such flexible loads scale their energy consumption up or down in response to supply, are spatially located near the site of overproduction, and absorb significant extra energy.

Data centers represent a potential power sink.
The amount of compute being performed can be scaled up or down with relative ease; data centers can be located anywhere and can be distributed geographically. 

American data centers consume approximately 70\,TWh yearly, 1.8\% of national energy consumption~\cite{enreport}. With estimates of opportunity power in CAISO and MISO sitting at 7--20\,TWh available per year~\cite{miso, caiso}, opportunity power in just these two regions has the potential to provide between 10-30\% of the power needed by data centers.

\subsection{A server's midlife crisis}

Zero-carbon computing also requires decarbonizing the embodied energy of that computation.
Embodied energy includes energy for mining, refining, and manufacturing, which can be quite significant~\cite{emergy11}.
Unfortunately, these processes may result in the majority of the lifetime CO$_2$ emissions associated with a device~\cite{c02-server}.

While very old servers can be acquired for cheap, many servers are in a ``midlife crisis.'' Their embodied energy would be wasted if recycled for their raw materials, but they are too old to deploy in most high-end data centers.
Therefore, a zero-carbon cloud will need to make use, as much as possible, of
existing computing equipment, notably equipment that has been replaced after
hardware refresh and would have ended up as eWaste~\cite{chien-extendinglife}.

Some in industry have sought to squeeze out  squeezing every embodied Joule out of using these ``midlife'' compute platforms by using them to run bulk compute tasks, such as training machine learning models~\cite{lancium}. However, in this work, we argue that these machines can also benefit general cloud compute tasks, namely the common cloud tasks that do not saturate every CPU core in a server.  Indeed, it  has proved difficult to saturate all available CPU cores with typical cloud workloads~\cite{carbyne,decima}.%

\begin{figure}
    \centering
    \includegraphics[width=\columnwidth]{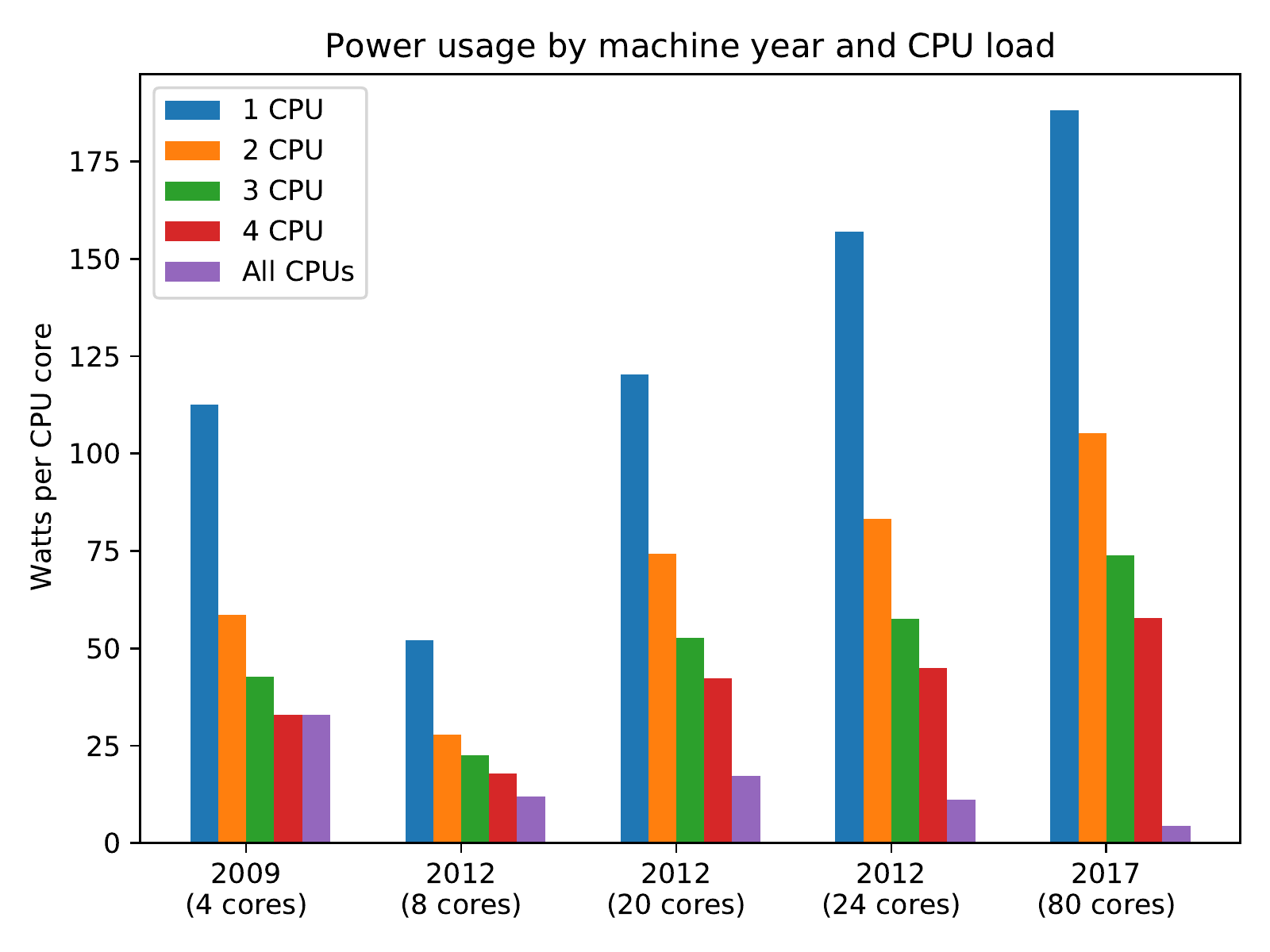}
    \vskip -1em
    \caption{Ratio of server power usage to 100\% utilized core count. Three 2012 server configurations are shown.}
    \label{fig:watts_per_core}
    \vskip -1em
\end{figure}

For ``midlife'' servers at reduced CPU load for zero-carbon cloud tasks, we compared the watts per number of active CPU cores for servers manufactured from 2009 to 2017 (\cref{fig:watts_per_core}).
The average power per core when using only a few cores has not changed significantly over time, likely due to static thermal limitations in processor packages. Thus ``midlife crisis'' servers that have been decommissioned from high-performance workloads can run more dynamic jobs on a zero-carbon cloud platform with similar operating energy as more modern servers, without incurring any additional embodied energy.

\section{\name design}

Abundant renewable energy and discarded hardware are available and can be used to build low-carbon computing. However, there are many design challenges faced in building \name as a truly zero-carbon platform. Here we describe those challenges, and point to potential solutions.

\subsection{\containername: A modular building block}
\label{sec:systems:design}

Our compute clusters must be physically located in regions with substantial renewable energy. We encapsulate compute, storage, and networking close to these power sources in a component we call a \containername. The Sun's and Google's modular data center architectures are likely models, due to their portable and weatherproof design.

Within each \containername we imagine mostly legacy equipment with a minimal addition of new equipment. Legacy equipment has a zero carbon footprint and low or zero capital cost, whereas newer equipment has a high upfront cost but lower operational cost due to density and power efficiency. Thus, what mixture of old and new equipment maximizes energy efficiency while minimizing overall cost and footprint? As described below, we assume a fiber interconnect to the Internet. The ``core'' of a \containername's network should be built from new, high-speed networking gear (in 2021 this would likely be 100\,Gbps switches), with long-reach optical transceivers facing the outer world. Each rack should have a comparable (e.g., 100\,Gbps) ToR switch, with short-reach optics between racks and optics or copper to servers within a rack.

Although deploying entirely legacy server equipment would maximize cost savings and minimize the embodied carbon/energy aspect of the \containername, modern data-intensive workloads are almost always memory constrained, rather than CPU constrained according to data published by Alibaba in 2019~\cite{alibaba}, where memory was fully utilized over 80\% of the time, and often nearly all the time, while at the same time the average CPU load was under 40\%, peaking at under 70\%.

We propose a \emph{hybrid} deployment of servers within each \containername: the majority consisting of older machines (with less cores and memory) as well as a small number of modern RAM-dense machines. Today machines with several terabytes of RAM are commercially available, and through techniques such as network disaggregation~\cite{gao-disag,porterdisag} and network-enabled swapping~\cite{infiniswap}, older machines could ``spill'' some of their working set across the network to these RAM machines. Gao et al.~\cite{gao-disag} study the feasibility of applying this idea to data-intensive processing jobs. A single RAM-dense machine with one or two 100\,Gbps NICs could interconnect anywhere from 4 to 20 legacy servers at sustained bandwidths of 10- to 25-Gbps. In addition to RAM-dense modern servers, a small number of servers that are dense with flash storage could logically connect to the \containername's high-speed network core. Intel has a 1U chassis that can potentially host up to 1 petabyte of m.2 flash storage modules~\cite{intel-ruler-formfactor}.

Thus, as workloads have increased their memory demands at a rate faster than their compute demands, configuring a hybrid design with a large number of older ``compute'' servers coupled with a small number of newer RAM- and Flash-storage nodes has the potential to support the widest possible set of workloads at the lowest monetary- and embodied energy-costs.

\subsection{Execution contexts}

Unlike stable data center environments that can offer long-running VMs to end users, \containernames have to respond to increases and decreases of demand, both on short timescales and potentially with little warning. Thus we argue that the overhead of setting up, deploying, and migrating VMs might be too heavyweight for rapidly changing environmental conditions. Instead, the recent development of FaaS (Function-as-a-service) and ``serverless'' computing better fits the timescales and statelessness we target.

A range of applications have been ported to serverless platforms, including video compression~\cite{excamera}, video processing~\cite{sprocket}, and data analysis~\cite{pywren}. Common among these applications is the ability to rapidly \emph{burst} the level of parallelism to match available compute or energy resources. Backing the activation of these functions are container images and requisite datasets for processing. These inputs could be stored on the flash-dense infrastructure nodes described above. This permits them high-speed access to the legacy servers and to the high-speed core for when entire jobs need to be migrated.

\subsection{Moving data and computation}

\begin{figure}[t]
    \centering
    \includegraphics[width=\columnwidth]{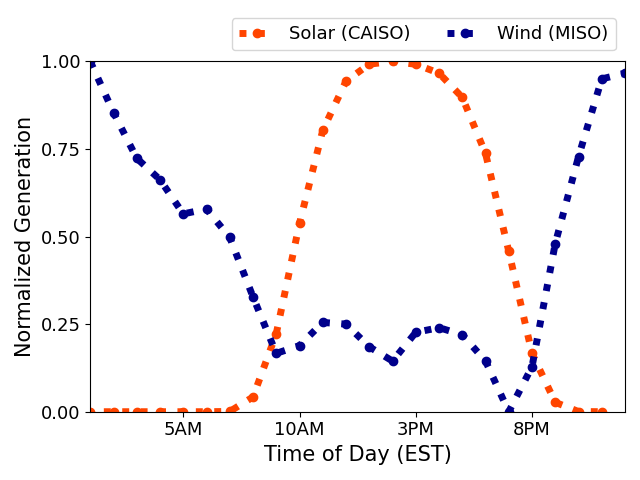}
    \vskip -1em
    \caption{Hourly renewable generation for CAISO (solar-dominant) and MISO (wind-dominant). Wind and solar production tend to be out of phase.}
    \label{fig:cm_prices}
    \vskip -1em
\end{figure}

\Cref{fig:cm_prices} shows how renewable power production changes over time in five-minute intervals. Each region does not always have curtailed power ready. Also, the regions are not in phase with one another.
Because each region has ephemeral availability of curtailed power, we must support dynamically relocating customer data and workloads to other locations when a site's available power changes. VM migration is one response, but we discuss in this section that our needs for migration are different than that of a traditional cloud platform.
We argue that new reliability metrics must be created and communicated to users in 
\name; we hypothesize that we may face the reality of fluctuating power generation by dynamically shifting workloads from one location to another. 

\paragraph{For reliability.}
Reliability is key for a cloud platform. Users expect that containers or VMs will meet SLOs without being affected by external factors like node and network failures.
We focus on platform reliability in two contexts: within a deployment site and between sites, each of which must meet different requirements.

Cloud platform reliability inside of a deployment site can build upon a wealth of prior work. VM migration is well studied and previous solutions can guarantee smooth site operation~\cite{andromeda,rocksteady,vsphere}. Leveraging the state replication and fault tolerance techniques described in such work is more than sufficient to ensure that any one deployment site can be maintained within the bounds of its own walls.

The more challenging and significant issue comes from how reliability is achieved as a whole between deployment sites. The core issues derive from the inalienable fact that the compute ability of a deployment site is non-static, and may even be taken completely offline. 

\paragraph{In response to power down/up events.}
We must ensure that when a deployment site's power availability changes, that we can proportionally shift the deployment load appropriately. More concretely, if site A loses renewable power while site B gains power, we need a way to transparently move customer jobs from site A to B. 

Previous work in VM migration has considered movement of a single VM, which typically takes \textasciitilde{}10's of milliseconds~\cite{andromeda}. However, in our case, we must understand how to move a set of VMs across a WAN without a significant impact to the user over the course of minutes.

To shift users from one region to another, we must know the job dependency graph to detect what data and services must be updated between regions. Detecting the DAG in cloud environments is a well-studied problem, with recent work~\cite{alibaba_dag} revealing that these DAGs have predictable and detectable forms. Modern cloud job schedulers~\cite{carbyne,decima} have used DAG information to provide insight to efficiently schedule large sets of jobs in cloud datacenters.

Using power production data similar to that in \cref{fig:cm_prices}, we can predict how available power will change over time. We similarly can predict how long it will take to move VMs and user data from one region to another. Using these two metrics along with DAG analysis and scheduling, it becomes possible to freeze and move VMs/data early enough before the region becomes unavailable to do so.

However, meeting homogeneous VM needs is challenging. We discuss in \cref{sec:systems:design} that we plan to deploy older hardware at each location. Even if site B has more power, it may not have an equivalent amount of compute relative to that power. Site B may also not have any curtailed power, as shown by the simultaneous fluctuations in price in \cref{fig:cm_prices}.
Thus, it may become necessary to freeze jobs and write them to cold storage and resume them when power returns. For those who pay more, jobs could be moved between ISOs to follow renewable capacity. 

\subsection{Inter- and intra-container networking}

Traditional data center networking assumes an always-on model. Such fat-tree deployments assume all-or-nothing power availability. \containernames do not have constant power, so we must scale the network to both take advantage of opportunity power and to scale back when power is scarce.

A fat-tree model still provides significant advantages for intra-container networking. ElasticTree~\cite{elastictree} provides a way to flexibly enable and disable parts of a fat-tree network based on job demand, which we can extend to scale the network based on available power.

Inter-datacenter networking requires high bandwidth, always-available links to move data between regions.\footnote{Independently, we must make the Internet zero-carbon~\cite{intermittent12}.}
Such links are likely to begin with wide-area fiber links already laid in parallel with construction of renewable energy generators~\cite{abb_fiber}. The control and network of these links depends on the specifics of the deployment location.

It is possible that dedicated links between locations can be provided, which will be used as necessary to shift jobs between locations as discussed previously. If no links are available, it remains as an open question whether links through ISP exchanges can provide sufficient spot bandwidth to transfer data and jobs between regions.

\subsection{Renewable energy and carbon SLOs}
We must carefully track carbon emissions due to each individual job. Thus we require a fine-grained energy and carbon accounting system in software and hardware, to track the energy use ``billable'' to each task and customer.

Given the physical hardware of a specific \containername, we must include its embodied energy, new and old alike. Performing such calculations is known to be challenging and error prone, as there is often scarce data on life-cycle energy for hardware, energy-accounting system boundaries are inherently fuzzy, and amortization is subjective. We do not expect to innovate on this matter but just to build upon the best available data for the hardware in use.

\subsection{Grids and Pricing}
Many organizations have begun to track their carbon footprints. Our aim with \name is to deliver near-zero carbon computation. To this end, users can specify their target renewable energy usage (e.g., ``no less than X\% renewable'') and/or carbon targets (e.g., ``no more than Y kg of CO$_2$e''), which, given \name's goals, are likely always to be achievable given job completion flexibility.

Users of \name are also likely to be price sensitive. Thus we must expose real-time power price information from the relevant ISO(s) and an estimate of ongoing fixed costs (e.g., cooling, bandwidth). 
Both CAISO and MISO report real-time, regional prices~\cite{miso_site,caiso_site}. Furthermore, since prices follow patterns, price prediction is feasible.

\subsection{\containername siting}
Siting of \containernames must be grid specific.
For those with large-scale renewable deployments far from loads (e.g., wind farms in rural areas), we would likely locate more \containernames in a higher density near generation sources.
By focusing on these high-production sites first, we can get large gains at relatively low cost: MISO's top producing wind farms each had as much as 250\,MW of opportunity power available at duty factors of up to 70\%~\cite{miso}.
In other areas with more diffuse renewable sources, we would scatter \containernames across the region.
This overall approach aims to ensure that opportunity power will be available at some subset of \containernames at any time.

\section{Conclusion}
Anthropogenic climate change, caused in large part by fossil-fuel based power generation and manufacturing processes, poses an existential threat to our civilization.
Over the next decade, with solar and wind installed prices already below that of virtually all fossil fuels, we may face a glut of renewable electricity yet will struggle to use it.
Computing has a high electricity and embodied footprint, yet its flexible nature across place, time, and type of equipment lets us mitigate its climate impact. We argue for the creation of large-scale compute platforms that can use renewable generation and older hardware to sustain sustainable computing for the next generation.

\bibliographystyle{plain}
\small
\bibliography{main}

\end{document}